\documentclass[conference]{IEEEtran}
\IEEEoverridecommandlockouts
\usepackage{cite}
\usepackage{amsmath,amssymb,amsfonts}
\usepackage{graphicx}
\usepackage{textcomp}
\usepackage{xcolor}
\usepackage{algpseudocode}
\usepackage{mathtools} 
\usepackage{amsmath}
\usepackage{amssymb}
\usepackage{amsthm}
\usepackage{mathtools}
\usepackage[ruled, vlined, linesnumbered]{algorithm2e}
\usepackage{color,soul}

\usepackage{cite}
\usepackage{subfigure}
\usepackage{stmaryrd}
\usepackage{breqn}
\usepackage{stackengine}
\usepackage{verbatim} 
\usepackage{glossaries}
\usepackage{color,soul}
\usepackage{xcolor}
\def\BibTeX{{\rm B\kern-.05em{\sc i\kern-.025em b}\kern-.08em
    T\kern-.1667em\lower.7ex\hbox{E}\kern-.125emX}}
\begin{document}

\title{Optimizing Downlink C-NOMA Transmission with Movable Antennas: A DDPG-based Approach\\
%\author{Author 1, Author 2, Author 3 and Author 4}
%Cooperative Sensing in Rate Splitting Enabled ISAC Systems
}
\author{Ali Amhaz, Mohamed Elhattab, Chadi Assi and Sanaa Sharafeddine}
%\and
%\IEEEauthorblockN{Mohamed ElHattab}
%\thanks{\noindent This work was supported in part by Concordia University, and the Natural Sciences and Engineering Research Council of Canada (NSERC).}
%\and
%\IEEEauthorblockN{Chadi Assi}
%
%\and
%\IEEEauthorblockN{Sanaa Sharafeddine}
%}
%\author{Ali Amhaz, Mohamad Elhattab, Sanaa Sharafeddine, and Chadi Assi  \thanks{A. Amhaz, M. Elhattab and C. Assi are with Concordia University, Montreal, Quebec, H3G 1M8, Canada (email: ali.amhaz02@lau.edu, mohelhattab@gmail.com, assi@mail.concordia.ca).}
%\thanks{Sanaa Sharafeddine is with the Department of Computer Science, American University of Beirut, Beirut 1107 2020, Lebanon (e-mail: ss30@aub.edu.lb).}}

\maketitle
\begin{abstract} 
This paper analyzes a downlink C-NOMA scenario where a base station (BS) is deployed to serve a pair of users equipped with movable antenna (MA) technology.
The user with better channel conditions with the BS will be able to transmit the signal to the other user providing an extra transmission resource and enhancing performance. Both users are equipped with a receiving MA each and a transmitting MA for the relaying user. In this regard, we formulate an optimization problem with the objective of maximizing the achievable sum rate by jointly determining the beamforming vector at the BS, the transmit power at the device and the positions of the MAs while meeting the quality of service (QoS) constraints. Due to the non-convex structure of the formulated problem and the randomness in the channels we adopt a deep deterministic policy gradient (DDPG) approach, a reinforcement learning (RL) algorithm capable of dealing with continuous state and action spaces. Numerical results demonstrate the superiority of the presented model compared to the other benchmark schemes showing gains reaching 45\% compared to the NOMA enabled MA scheme and 60\% compared to C-NOMA model with fixed antennas. The solution approach showed   93\% accuracy compared to the optimal solution. 
\end{abstract}
 
\section{Introduction}
The future of wireless networks, especially 6G, will encounter an unprecedented level of user density driven by the huge growth in the connected devices and surging wireless traffic. This trend aligns with the rise of the Internet of Everything (IoE) that is anticipated to serve billions of users revealing the need for higher data rates, ubiquitous connectivity, and ultra-low latency communication \cite{9123680}.  With the aim of satisfying these demands, researchers already started investigating the suitable multiple access techniques and effective network architectures. Multiple-input multiple-output (MIMO), a main technology in 5G networks, represented an effective approach to improve the degrees of freedom (DoFs) yielding better efficiency in wireless communications \cite{6798744}. Moreover, such a technique showed performance gains when combined with other technologies, i.e., reflective intelligent surfaces (RIS), coordinated multi-point (CoMP), integrated sensing and communication (ISAC), and others \cite{10584278, 9110912}. Nonetheless, MIMO relies on a fixed architecture for the different antennas losing the chance to further exploit the spatial diversity potentials \cite{10318061}. 
\par In light of this, several studies recently started focusing on the movable antenna (MA) technology to better harness the DoFs in the spatial domain. This technology operates by adjusting the position of the antenna spatially to enhance the channel quality representing a more clever approach than antenna selection (AS) technique that requires a huge number of antennas and comes with additional implementation costs \cite{1284943}. With as few as a single antenna, MA technology, can maneuver the antenna in a continuous manner improving spatial diversity and combating severe fading scenarios \cite{zhu2024movableantennaenhancedmultiusercommunication}. In \cite{10416896}, the authors revealed the advantage of MA technology in achieving reduced power consumption by adjusting the position of the MA at the communication users.  

Another key component for boosting the performance of communication systems is the wise selection of multiple access technologies. Non-orthogonal multiple access (NOMA) proved to attain significant gains, in terms of spectral and energy efficiency, over traditional orthogonal approaches benefiting from the ability to serve multiple users on the same spectrum and exploiting successive interference cancellation (SIC) techniques \cite{10505024}. However, NOMA experiences a decline in performance when users are co-located, forcing the base station (BS) to allocate additional resources to those with poor channel conditions. As a solution, cooperative NOMA (C-NOMA) emerged as an improved version of NOMA which leverages the device-to-device (D2D) links between users, thus provides an extra transmission resource to the far users and improves the overall performance. In \cite{10505024}, the authors analyzed the performance of  user-assisted C-NOMA system with the objective of maximizing the achievable sum rate. Numerical results proved its superiority over traditional benchmarks including NOMA.  

The first attempt to integrate NOMA and MA technologies was in \cite{zhou2024movableantennaempowereddownlink} where the authors formulated an optimization problem to maximize the achievable sum rate of the communication users by determining the power allocation factor and the MA positions. However, as stated above, NOMA technology still suffers from reduced performance in the case of coexistence of near and far users because of the co-channel interference and the need for the BS to assign high power to the far users. Moreover, \cite{zhou2024movableantennaempowereddownlink} tackles the single antenna BS scenario making it hard to generalize it to practical use cases. In light of this, and to the best of our knowledge, the MA-empowered C-NOMA scheme has yet to be explored in the literature, making it an interesting area for further investigation. Therefore, in this work, we investigate the integration of C-NOMA with MA technology at the transmitting and receiving antennas of the users where the BS is equipped with multiple antennas. We formulate an optimization problem with the aim of maximizing the achievable rate by determining the beamforing vectors at the BS, the device transmit power, and the MA positions while respecting the quality of service (QoS) constraints. The formulated problem showed to be non-convex and difficult to solve because of the high coupling between the decision parameters and the randomness in the channels. In this regard, we adopt a deep deterministic policy gradient (DDPG) framework to solve it efficiently and effectively. Finally, numerical results demonstrated the advantage of our model over other benchmark schemes and the accuracy of the RL-based solution. 

\begin{figure}[!t]
    \centering    \includegraphics[width=1\columnwidth]{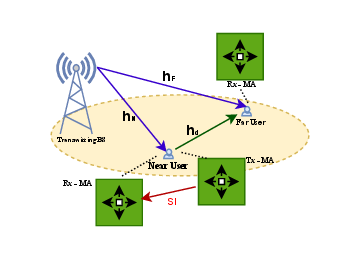}
    \caption{System model.}
	\label{fig_11}
\end{figure}

\section{System Model}
In this paper, we study the scenario of a downlink C-NOMA network where MA technology is deployed at the communication users. It consists of a BS equipped with N-antennas and a set of 2 communication users as depicted in Fig. 1. The user experiencing better channel conditions with the BS is referred to as near user $N$, on the other hand, the second user is referred to as far user $F$. Both $N$ and $F$ support a single antenna MA at the receiver side, while $N$ is also equipped with a single MA for the transmission. The BS will serve the 2 users using NOMA technology, i.e., transmits a superimposed signal to both users utilizing the same spectrum. Based on that, user $N$ starts first by decoding the signal of user $F$, then removes it using the successive interference cancellation (SIC) technique to be able to decode its own signal. Next, user $F$ can decode its signal by considering the signal of user $N$ as interference. In this study, the D2D link between user $N$ and $F$ is exploited in order to improve the signal of user $F$ benefiting from the fact that user $F$ signal is being decoded at user $N$ side and the ability to combine the signal from both sources using maximum ratio combining (MRC) technique, thus supporting C-NOMA scheme. In this model, we assume that user $N$ employs a decode and forward (DF) strategy. In addition, the transmissions of the BS and user $N$ occur at the same time following the full-duplex (FD) transmission mode resulting in a residual self-interference (SI). Here, it is worth noting that the the minimum distance between the transmitting and receiving MA at user $N$ is $\frac{\lambda}{2}$ to reduced the effect of SI.  

%the receiving antennas at both users $N$ and $F$, the divergence in the signal propagation from the m-th transmit path of the BS n-th antenna and the receiving antennas position (i.e., $r_N = [x_N, x_N]^T$ for user N and $r_F = [x_F, x_F]^T$ for user F) can be illustrated as

\subsection{The Channel Model}
In this model, we assume that the different channels follow a field response channel \cite{zhou2024movableantennaempowereddownlink}, \cite{10318061}. The channel between the BS and user $N$ is denoted as $\boldsymbol{h}_N$, the channel between the BS and user $F$ is expressed as $\boldsymbol{h}_F$, and the channel corresponding to the D2D channel between the near and far users is $h_d$. The BS origin is assumed to be $O^t_n = [o_n, o_n]^T$ for each antenna $n$ at the BS. For the receiving antenna at users $N$ and $F$, we assume that the origins are $O_{N, r} = [0, 0]^T$ and $O_F = [0, 0]^T$, respectively. On the other hand, the origin of the transmit antenna at user $N$ is denoted as $O_{N,t} = [0, 0]^T$. The number of transmit paths between each antenna of the BS to the receiving antenna of both users is $L_B$. Similarly, the number of paths from the transmit antenna of user $N$ to the receiving antenna of user $F$ is $L_A$. On the other hand, $L_R$ denotes the number of receive paths at both users. Accordingly, we define the  path response matrix (PRM) between the BS and the users $N$ and $F$ as $\Sigma_N$ and $\Sigma_F$, respectively. For the link between user $N$ and $F$ it is expressed as $\Sigma_{N,F}$ \cite{zhu2024movableantennaenhancedmultiusercommunication}. Following the model adopted in \cite{10318061}, the mobility in MA will not change the different angles in the model (i.e, angle-of-arrival (AoA) and angle-of-departure (AoD)) for the various paths in the different channels  $\boldsymbol{h}_N$, $\boldsymbol{h}_F$, and ${h}_d$. For both users $N$ and $F$, the divergence in the signal propagation for the m-th transmit path of the BS's n-th antenna (with the position $t_n = [x_n, y_n]^T$) and the origin can be illustrated as $\rho^t_{m,a} = x_n^t \cos \theta^t_{m,a} \sin \phi^t_{m, a} + y_n \sin \theta_{m,a} $, with $\theta_{m,a}$, $ \forall a \in \{N, F\}$, being the elevation of AoD and  $\phi_{m,a}$, $ \forall a \in \{N, F\}$, corresponding to the azimuth of AoD \cite{10318061}. Thus, we express the transmit field response vector (FRV) of the n-th antenna at the BS as:
\begin{align}
    &\textbf{g}_a(t_n) = [e^{-j \frac{2\pi}{\lambda} \rho^t_{1, a} (t_n)}, e^{-j \frac{2\pi}{\lambda} \rho^t_{2, a} (t_n)}, ..., e^{-j \frac{2\pi}{\lambda} \rho^t_{L_b, a} (t_n)} ], \\
    &\hspace{3cm}\forall a \in \{N,F\}, \nonumber
\end{align}

\noindent In the same way, we derive the FRV of both users at the receiving side as: 
\begin{align}
    &\textbf{f}_a(r_a) = [e^{-j \frac{2\pi}{\lambda} \rho^r_{1, a} (t_k)}, e^{-j \frac{2\pi}{\lambda} \rho^r_{2, a} (t_k)}, ..., e^{-j \frac{2\pi}{\lambda} \rho^r_{L_R, a} (t_k)} ], \\
    &\hspace{3cm}\forall a \in \{N,F\}, \nonumber
\end{align}
\noindent where the propagation divergence corresponding for the $d$-th receiving path between the position of MA ($r_a = [x_a^r, y_a^r]^T$) and its origin is $\rho^r_{d,a} = x_a^r \cos \theta^t_{d,a} \sin \phi^r_{d, a} + y_a^r \sin \theta_{d,a}$, $\forall a \in \{N,F\}$, with $\theta^t_{d,a}$ and $\phi^r_{d, a}$ being the elevation and azimuth AoA between the d-th path and the different users, respectively. Similarly, we derive the transmit and receive FRV corresponding to the link between the two users $N$ and $F$ and we denote it as $\boldsymbol{g}_N(t_d)$ and $\hat{\boldsymbol{f}}_F(r_F)$, respectively, with $t_d = [x_N^t, y_N^t]^T$ being the position of transmitting MA at user $N$. Finally, the channels are defined as:
\begin{align}
    &\boldsymbol{h}_N = \boldsymbol{f}_N^T (r_N). \boldsymbol{\Sigma}_N . \boldsymbol{G}_N, 
\end{align}
\begin{align}
    &\boldsymbol{h}_F = \boldsymbol{f}_F^T (r_F) . \boldsymbol{\Sigma}_F . \boldsymbol{G}_F, 
\end{align}
\begin{align}
    &h_d = \hat{\boldsymbol{f}}_{F}^T (r_F) . \boldsymbol{\Sigma}_{N, F} . \boldsymbol{g}_N(t_d),
\end{align}
where $\boldsymbol{G}_a = [\textbf{g}_a(t_1), \textbf{g}_a(t_2), ..., \textbf{g}_a(t_N) ] $ $\forall a \in \{N,F\}$ being the  transmit field response matrix (FRM) from the BS to the user $N$ and $F$. 

\subsection{The Communication Model}
Starting with the signal received by user $N$ from the BS, it can be represented as:
\begin{equation}
y_{N} =  \boldsymbol{h}_N^H \left(\boldsymbol{w}_Fs_F + \boldsymbol{w}_Ns_N\right)  + h_{SI} \sqrt{P_N}  s_F +  n_N, \label{eq1} 
\end{equation}
where the transmission power of user $N$ is denoted as $P_N$, and $n_N$ corresponds to the additive white Gaussian noise (AWGN) following $\mathcal{CN}(0, \sigma^2)$ with zero mean and variance $\sigma^2$. The first part of \eqref{eq1} is due to the transmission of the BS with $s_N$ and $s_F$ being the signals of user $N$ and user $F$, respectively. Also, $w_N$ and $w_F$ are the beamforming vectors for user $N$ and $F$ transmissions, respectively. On the other hand, the second part corresponds to the SI effect due to the FD transmission mode. Here, $h_{SI}$ denotes the SI channel and it adopts a complex symmetric Gaussian random variable with a zero mean and variance $\Omega_{SI}^2$. Meanwhile, user $F$ will receive its message from two sources (the BS and user $N$) in which the received signal can be expressed as follows:
\begin{equation}
y_{F} =  \boldsymbol{h}_F^H (\boldsymbol{w}_Fs_F + \boldsymbol{w}_Ns_N) + h_{d} \sqrt{P_N}  s_F +  n_F, \label{rx_fd} 
\end{equation}
where $n_F$ is the AWGN following $\mathcal{CN}(0, \sigma^2)$. User $N$ starts first by decoding the message that corresponds to user $F$. Hence, at user $N$, the signal-to-interference-plus-noise ratio (SINR) to decode the message of user $F$ at user $N$ can be expressed as:
\begin{equation}
\Lambda_{N \rightarrow F} =\frac{| \boldsymbol{h}_N^H \boldsymbol{w}_F|^2} {|\boldsymbol{h}_N^H \boldsymbol{w}_N |^2 + P_N |h_{SI}|^2 + \sigma^2},
\end{equation}
Hence, the achievable data rate can be expressed as 
\begin{equation}
    R_{N \rightarrow F} = \log_2(1 + \Lambda_{N \rightarrow F}).
\end{equation}
By leveraging the SIC process at user $N$, this decoded signal of user $F$ can be removed. Hence, user $N$ decodes its own signal with the SINR expressed as follows:
\begin{equation}
\Lambda_{N \rightarrow N} =\frac{|\boldsymbol{h}_N^H\boldsymbol{w}_N |^2} {P_N |h_{SI}|^2 + \sigma^2},
\end{equation}
and the rate expression is:
\begin{equation}
    R_{N \rightarrow N} = \log_2( 1 + \Lambda_{N \rightarrow N}).
\end{equation}
Due to the cooperative scheme, user $F$ will receive its signal from the BS and user $N$. Thus, it can exploit the MRC technique to combine the two received signals yielding the following SINR:
\begin{equation}
\Lambda_{MRC} = \frac{|\boldsymbol{h}_F^H \boldsymbol{w}_F |^2} {|\boldsymbol{h}_F^H \boldsymbol{w}_N |^2 + \sigma^2} + \frac{P_N |h_d|^2}{\sigma^2}.
\end{equation}
 Accordingly, the achievable data rate to decode the message of user $F$ after applying the MRC process can be denoted as:
 \begin{equation}
     R_{MRC} = \log_2(1 + \Lambda_{MRC}).
 \end{equation}
 Nonetheless, achieving the rate $R_{MRC}$ is restricted to the ability of user $N$ to decode user $F$'s signal. Thus, we express the final rate to decode the message of user $F$ as:
 \begin{equation}
     R_{F \rightarrow F} = \min ( R_{MRC}, R_{N \rightarrow F}).
 \end{equation}
\section{Problem Formulation}
With the aim of maximizing the sum rate of the communication users, we formulate the optimization problem to determine the beamforming vectors at the BS, the transmit power of user $N$, and the locations of the transmit and receiving MA at both users while guaranteeing the QoS constraints of the users. Hence, the formulated problem is expressed as:
%\max_{\boldsymbol{p}_F, \boldsymbol{p}_N ,P_{N}, \alpha}  \ \;   \rho_c(R_{N} +   R_{F}) + \rho_s P(\theta_m)
\allowdisplaybreaks
\begingroup
\begin{subequations}
\begin{align}
&\mathcal{P}_1: \max_{\boldsymbol{w}_F, \boldsymbol{w}_N ,P_{N}, t_d , r_N, r_F}  \ \;   (R_{N \rightarrow N} +   R_{F \rightarrow F}) \\
\textrm{s.t.}  \ \; 
     &\boldsymbol{w}_F^H \boldsymbol{w}_F + \boldsymbol{w}_N^H \boldsymbol{w}_N \leq P_T, \label{C_7} \\
   & 0 \le P_{N} \leq P_{NF}, \label{C_2}\\
   & R_{a \rightarrow a} \geq R_{th}, \qquad \forall a \in \ \{\mathrm{N,~F}\}\label{C_3}\\
   & R_{N \rightarrow F} \geq R_{th}, \label{C_5}\\
   & t_d \in \mathcal{D}_{t_d}, \label{tx1}\\
   & r_N \in \mathcal{D}_{r_1}, \label{rx1}\\
   & r_F \in \mathcal{D}_{r_2}, \label{rx2}
\end{align} 
\end{subequations}
\endgroup
where $P_{T}$ and $P_{NF}$ represent the power budgets at the BS and the user $N$, respectively. $\mathcal{D}_{t_d}$, $\mathcal{D}_{r_1}$, and $\mathcal{D}_{r_2}$ denote the mobility area of the different MAs and they can be expressed as $A \times A$ in a 2D space. Constraints \eqref{C_7} and \eqref{C_2} is to respect the power budgets at the BS and user $N$. Meanwhile, constraints \eqref{C_3} and \eqref{C_5} guarantee a minimum achievable rate at the communication users and a successful SIC process at user $N$, respectively. Finally, constraints \eqref{tx1}-\eqref{rx2} restrict the mobility regions at the transmit MA at user $N$, and the receiving MAs at users $N$ and $F$, respectively. 
\section{Solution Approach}
The optimization problem introduced in the previous section is non-convex due to the high coupling between the variables resulting in non-convex constraints and objective function. In this regard, and to handle the randomness in the different channels of the system model, we adopt a DDPG framework, a reinforcement learning approach to solve the optimization problem. Such an algorithm produces actions upon interacting with the environment by maximizing a reward function and applying penalties in the case of unsatisfied constraints.

In order to deal with the challenges accompanied with the continuous state and action spaces and to avoid the performance loss imposed by the discretization in the traditional reinforcement learning algorithms, DDPG framework was designed to deal with the aforementioned problems. Such an algorithm adopts the "Actor-Critic" concept in which the "Actor" component produces the policy (that defines the actions) in the environment. The "Critic" part on the other side is responsible of assessing the actions by the ability to generate the Q-function. The two components consist of a couple of neural networks, namely, training and target networks. The training networks for Actor and Critic parts are denoted as $Q$ and $\mu$, respectively. They learn from past experiences to adjust the weights with the aim of improving policy and value estimation. The target networks for Actor and Critic parts are denoted as $Q'$ and $\mu'$, respectively. Such networks assists in the stabilization process by producing target parameters for training networks. In this regard, we define the loss function that associates with this algorithm as \cite{9110869}:
\begin{align}
    &l(\theta) = \left(r^{(t+1)} + \eta \max_{a'} L'
    (\theta^{L'} | s^{(t+1)}, a')\right) - \nonumber \\
    &  L(\theta^{L} | s^{(t)}, a^{(t)}),
    \label{loss}
\end{align}
where the type of the networks are denoted by $H \in \{Q,\mu\}$  and $H' \in \{Q',\mu'\}$, $r^{(t+1)}$ represent the reward at the time step $t+1$, $\eta$ corresponds to the discount factor of the reward, $a^{(t)}$ is the action, and  $\theta^{H}$ and $\theta^{H'}$ denote the networks' weights. To update the weights of the training and target networks we adopt the same methods as in \cite{9110869} where the full details are omitted for brevity. Finally, $\tau_V$ ($V \in  \{Q,\mu\, Q',\mu'\}$) denotes the learning rate for the different actor "Actor-Critic" networks and it is utilized in \textbf{Algorithm 1}.  

\subsection{The Algorithm Framework}
\begin{algorithm}[!t]
\label{Algo_2}
\DontPrintSemicolon
\small
\SetAlgoCaptionLayout{centerline}
{\caption{\small Proposed DDPG Algorithm}}
  \textbf{Initialize}: 
  \SetAlgoCaptionLayout{centerline} $Q$ and $\mu$ networks with random weights $\theta^Q$ and $\theta^\mu$;
  
  \textbf{Initialize}:   $Q'$ and $\mu'$ networks with $\theta^{Q'} \leftarrow \theta^Q$, $\theta^{\mu'} \leftarrow \theta^\mu$
  
  \textbf{Initialize}:  replay buffer $BF$
  
\For{episode $= 1, M$}{
  Initialize environment state $s_1$;\;
  
   \For{$t = 1, T$}{
    Select action $a^{(t)}$ from $\mu$ ;\;

    Apply normalization for the beamformers;\;
    
    Execute $a^{(t)}$ and check $r^{(t+1)}$ and $s^{(t+1)}$;\;
    
    Place experience $(s^{(t)}, a^{(t)}, r^{(t+1)}, s^{(t+1)})$ in $BF$;\;
    
    Calculate Q value from Q network;\;
    
    Sample a random minibatch from $BF$;\;
    
    Minimize \eqref{loss} by updating critic Q;\; 
    
    Using sampled policy gradient, update network $\mu$;\;
    
    For $\mu'$ and $Q'$ networks, perform soft update: $\theta^{Q'} \leftarrow \tau \theta^Q + (1 - \tau) \theta^{Q'}$, $\theta^{\mu'} \leftarrow \tau \theta^\mu + (1 - \tau) \theta^{\mu'}$;\;}}
\end{algorithm}

\begin{figure*}
  \centering
  \subfigure[Convergence of the solution approach.]{\includegraphics[width=0.31\linewidth]{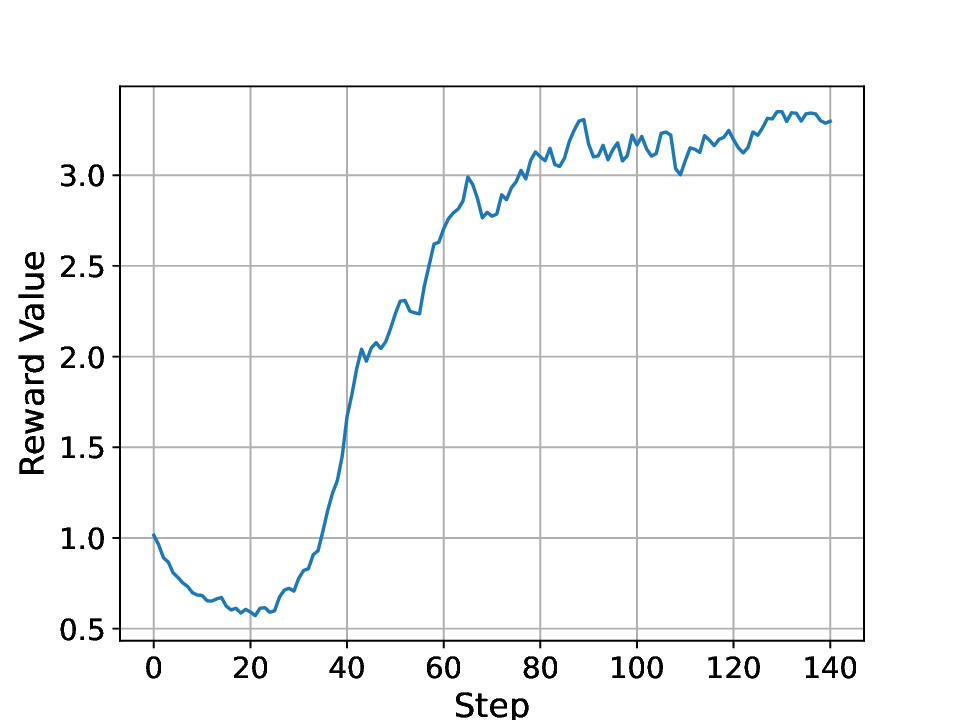}}
  \subfigure[BS power vs achievable sum rate.]
  {\includegraphics[width=0.31\linewidth]{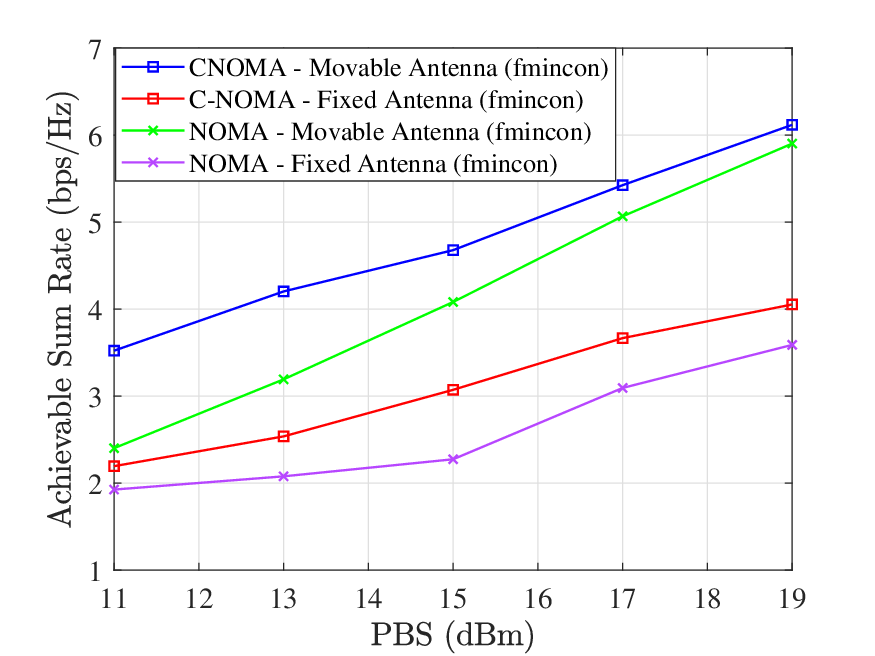}}
  \subfigure[Normalized mobility regoin versus the achievable sum rate.]{\includegraphics[width=0.31\linewidth]{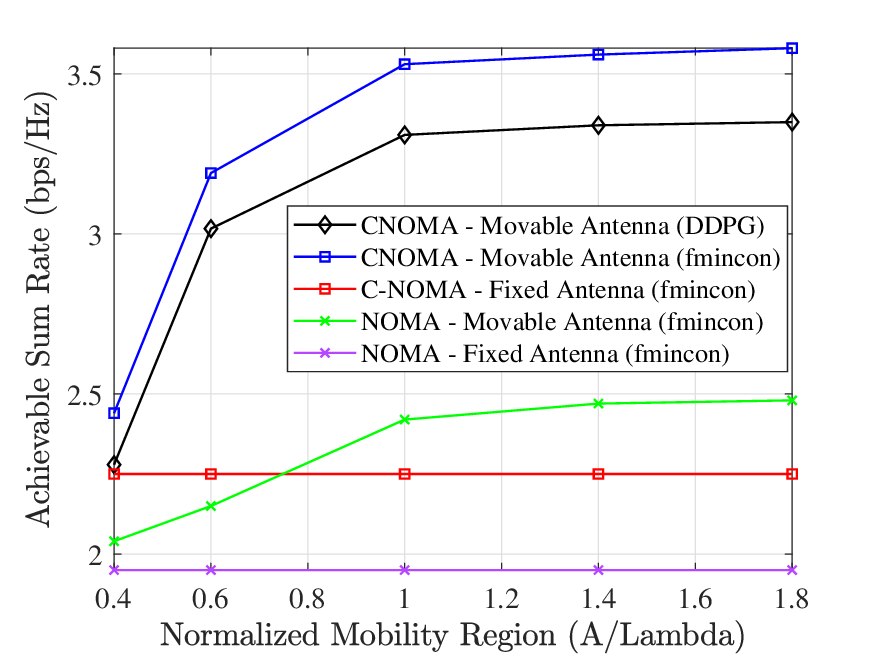}}
  \setlength{\belowcaptionskip}{-12pt} 
  \caption{Numerical results for the model under different parameters.}
  \label{three_figures}
  \vspace{-0.3cm}
\end{figure*}

The action, state, and reward functions which define the environmental properties are expressed as follows: 
\begin{itemize}
  \item  $a^{(t)} = \left[\boldsymbol{w}_1^{(t)}, \boldsymbol{w}_2^{(t)},P_n^{(t)}, x_{N,t}^{(t)} ,y_{N,t}^{(t)}, x_{N,r}^{(t)} ,y_{N,r}^{(t)}, x_{F,r}^{(t)} ,y_{F,r}^{(t)} \right]$ is the action consisting of the beamformers, power at near user and MA positions.  
  \item  $s^{(t)} = \left[ a^{(t-1)}, ||\boldsymbol{w}_1^{(t -1)}||^2,||\boldsymbol{w}_2^{(t-1)}||^2 \right. , P_N^{(t-1)},$ $ , \boldsymbol{h}_1, \boldsymbol{h}_2, \boldsymbol{h}_{N,F}]$ defines the state in the environment. Each state is formed of the previous action, different channels in the model, the previous power of user N, and the squared norm of the beamforming vectors. 
  \item $r^{(t)} = [R_N + R_F - (I_1(\min(R_N,  R_{N\rightarrow F},$$ R_{F}) - R_{th})*pen) - (I_2(x_{N,t} ,y_{N,t}, x_{N,r} ,y_{N,r}, x_{F,r} ,y_{F,r} )*pen)- (I_3(P_N )*pen)]$ represent the reward function with $R_N + R_F$ to ensure maximizing the objective function of our optimization problem. 
  \end{itemize}
 We utilize the function $I_1(.)$ in order to guarantee satisfying the QoS constraints in the optimization problem by imposing a penalty (with value $pen$) on the reward function. This function is defined as follow:
    \begin{equation}
I_1(x)  = \begin{cases}0, & \mbox{if } \mbox{x $\geq$ 0}, \\  1, & \mbox{if } \mbox{x $<$ 0}, \end{cases}
\end{equation}
$I_2(.)$ is to penalize the reward when MAs position is selected to be outside the designated area, and the $I_3 (.)$ is to impose a penalty when user $N$ power bedget is exceeded. 
%and the indicator function $I_2(.)$ is to penalize the reward function in the case of violating the UAV location constraint \eqref{e2} and \eqref{e22} and it is represented as follows:
%    \begin{equation}
%I_2(x)  = \begin{cases}0, & \mbox{if } \mbox{0 $\geq$ x $\geq$ $x_{max}$}, \\  1, & \mbox{otherwise}, \end{cases}
%\end{equation}
\noindent The penalty mechanism that was leveraged in the reward function handles the different constraints except the power budget constraint at the BS. In this regard, we rely of the normalization approach that was used by \cite{9110869} in order to prohibit exceeding the available power budget. Hence, the update mechanism of the beamforming vectors is expressed as: 
\begin{align}
&\boldsymbol{w}_i^{*(n)} = \boldsymbol{w}_i^{(n)}\sqrt{\frac{P_T}{P_t^{(n)}}},
\label{norma}
\end{align}
where $P_t^{(n)}$ can be defined as:
\begin{align}
&P_t^{(n)} = \sum^K_{i=1} || w_i^{(n)}||^2.
\end{align}
\subsection{Algorithm} %try making it better
We illustrate the pseudo code of our approach in Algorithm 1. The first step is to initialize the different neural networks for the "Actor-Critic" components along with the replay buffer BF that is responsible of storing experiences. The algorithm consists of a set of episodes in which the environment (i.e., channels) is reinitialized. Each episode involves a set of steps (total number is T) and within each one the action "a" is selected by utilizing the "Actor" component. Next, the beamforming normalization mechanism explained in \eqref{norma} is applied. Then it executes "a" and stores in BF the corresponding experience $(s^{(t)}, a^{(t)}, r^{(t+1)}, s^{(t+1)})$. From BF, a batch is sampled and the different network parameters are updated. Finally, the produced model is saved to be utilized in the testing phase.

\section{Numerical Evaluation}

\begin{table}[t]
\caption{Simulation Parameters}
\centering
\begin{tabular}{|l|c|l|c|}
    \hline
        Parameters & Values & Parameters & Values \\ \hline
        ~$\lambda$ & ~0.01 & ~$\xi$ & ~0.99 \\ \hline
        ~$L_A$ & ~6 & ~$\sigma^2$ & -100 dBm/Hz \\ \hline
        ~$A$ & ~2$\lambda$ & ~$L_{R}$ & ~6 \\ \hline
        ~$N$ & ~4 & ~$L_{B}$ & ~6 \\ \hline
        ~$R_{th}$ & ~0.7bps/Hz & ~$\alpha$ & ~3.9\\ \hline
        ~$\tau_Q$ & 0.001 & ~$\tau_{Q'}$ & ~0.001\\ \hline
        ~$\tau_\mu$ & ~0.001 & ~$\tau_{\mu'}$ & ~0.001\\ \hline
  \end{tabular} 
  \label{T1}
\end{table}

\begin{table}[t]
\caption{BS Power vs DDPG Algorithm}
\centering
\begin{tabular}{||c|c|c||}
    \hline
        BS Power (dBm) & DDPG Algorithm (bps/Hz) & Fmincon (bps/Hz) \\ \hline
        ~$11$ & ~3.309 & ~3.522  \\ \hline
        ~$13$ & ~4.016 & ~4.203  \\ \hline
        ~$15$ & ~4.402 & ~4.678  \\ \hline
        ~$17$ & ~5.145 & ~5.425  \\ \hline
        ~$18$ & ~5.900 & ~6.118 \\ \hline
  \end{tabular} 
\label{T2}
\end{table}
In this section, we assess the performance of the presented MA-CNOMA model by showing the corresponding numerical analysis. We illustrate the used parameters in Table 1. We averaged the generated results with over 200 realizations. The model is compared with a set of benchmarks defined below:
\begin{itemize}
  \item \textbf{MA-NOMA}: where 2 communication users (equipped with receiving MA) are served using NOMA technology.
  \item \textbf{F-C-NOMA}: where the C-NOMA technology is leveraged to serve users with fixed transmit and receiving antennas.
  \item  \textbf{F-NOMA}: where traditional NOMA is used to serve two users with fixed antennas.
  \end{itemize}

We examine in Fig 2 (a) the convergence of the DDPG algorithm by plotting the average reward at the end of every episode. Between episodes 0 and 20, we can observe that the average reward value tends to decrease followed by a steady increase. This is explained by the ability of the model to learn from experiences and maximizing the sum rate which is the objective of the optimization problem. Then, the value of the reward keeps on increasing until stabilizing around 3.5 proving the convergence of the algorithm.  

Fig. 2 (b) presents the achievable sum rate (bps/Hz) versus the power of the BS (dBm) for the different benchmarks using the fmincon tool (a MATLAB solver \cite{9586734}). It is clear that with the increase in the power budget of the BS, the achievable rate of the different schemes increases steadily with a superiority for the presented MA-CNOMA approach. This can be explained from one side by the advantage of the D2D link in assisting the user thus granting the BS additional freedom in maximizing the sum rate of the user $N$. On the other hand, the MA technology that is leveraged at the transmitting and receiving antennas of users $N$ and $F$ grants the ability to combat the fading effect by the flexible positioning of the antennas. Moreover, we can observe that with the increase in the power of the BS, the MA-NOMA and MA-CNOMA schemes approach each other which is interpreted by the abundance in the resources at the BS thus eliminating the necessity of relying on the D2D link. Another interesting observation is the ability of MA-NOMA scheme to outperform F-C-NOMA highlighting further the importance of this technology by exceeding the benefits provided by the D2D link. Finally, in order to evaluate the accuracy of our DDPG approach, we compare in Table II our approach versus the fmincon solution which provides the optimal solution. It is clear that the RL algorithm is producing close results with an accuracy reaching 93\% demonstrating its effectiveness. 

Fig. 2 (c) studies the effect of the mobility region area of the different MAs on the achievable sum rate for the different schemes. With the increase in the mobility region, the achievable rates for the MA-CNOMA and MA-NOMA scheme increase until the normalized mobility region reaches a value of 1. This can be interpreted by the higher flexibility for MA positioning which leads to better ability in suppressing the negative effect of fading. On the other hand, after the mobility region of value 1, the achievable sum rate tends to stabilize for the two schemes supporting MA technology, signifying the effective area of movement and the inability to further improve the rate. This is explained by the  MA reaching its maximum capacity of combating fading at this normalized mobility region value.

\section{Conclusion}
To conclude, this paper investigates a downlink C-NOMA mode where a base station (BS) serves two users utilizing movable antenna (MA) technology. The user with a stronger connection to the BS helps relay signals to the other user, offering an improved performance. Both users are equipped with receiving MAs, while the stronger user also has a transmitting MA to forward the far user’s message. We address this scenario by formulating an optimization problem aimed at maximizing the total achievable sum rate, optimizing beamforming at the BS, transmission power, and the position of the MAs, while respecting to the quality of service (QoS) requirements. Due to the complexity of the problem and the randomness in channels, a deep deterministic policy gradient (DDPG) reinforcement learning algorithm was applied to tackle the continuous state and action spaces. The numerical results confirm the proposed approach's superiority over existing benchmark methods.
\bibliographystyle{IEEEtran}
\bibliography{ref}

\end{document}